# On Delay Faults Affecting I/O Blocks of an SRAM-Based FPGA Due to Ionizing Radiations

Fatima Zahra Tazi[1], Claude Thibeault[1], Yvon Savaria[2], Simon Pichette[1], Yves Audet[2],

1Ecole de technologie supérieure, Montréal, QC, Canada
2Ecole Polytechnique de Montréal, Montréal, QC, Canada

*Abstract*

*Experimental means to characterize delay faults induced by bit flips and SEUs in I/O blocks of SRAM-Based FPGAs are proposed. A delay fault of 6.2ns sensitized by an events chain is reported.*

**Presenter:**
Fatima Zahra Tazi
PhD Student
Electrical Engineering Department
École de technologie supérieure
1100 Notre-Dame Street West,
Montreal, QC, Canada, H3C 1K3
Email: fatima-zahra.tazi.1@ens.etsmtl.ca

## I. INTRODUCTION

Despite the sensitivity to radiations of SRAM-based FPGAs, they have become increasingly popular among designers of avionics and even of space embedded systems. This growing interest is due mainly to their lower non-recurring engineering costs [5] as compared to ASICs, and their lower costs when compared to rad-hard FPGAs. The interest for state-of-the-art FPGAs also stems from the fact that they are usually fabricated with a technology a few nodes ahead [6] when compared to ASICs and to rad-hard FPGAs. This mitigates possible speed and density disadvantages that conventional FPGAs could have over ASICs and it translates into higher capability and speeds when compared to rad-hard FPGAs.

SRAM-based FPGAs contain Input/Output Blocks (IOBs), memory modules, as well as logic and routing resources, which are controlled by SRAM cells holding so-called configuration bits. SRAM-based FPGA sensitivity to radiations has been extensively studied over the years [7-14], with a particular attention dedicated to their configuration bits. The most commonly considered impacts of a bit-flip on these bits are opens and shorts. Recently [1], it was shown that delay faults could also be induced by such bit flips. So far, most of the research efforts have been dedicated to configuration bits controlling the FPGA core; giving less attention to the configuration bits controlling the IOBs. Very few researchers have investigated the behavior of the IOBs in the presence of SEUs [15, 16, 18], even if the possible effects, ranging from a slight power consumption increase to system damage, have been known for a while [18]. In [15], the impact of single event upsets (SEUs) on IOBs was emulated using a test bed allowing modifying the configuration bitstream of a Xilinx Virtex (XCV1000) FPGA. The logical values on the outputs were compared with those on a second Virtex used as a reference. They were specifically interested by Single Bit Upsets (SBUs) or Multiple Bits Upsets (MBUs) causing an IOB configured as an input to become an output. They observed that some configuration bits could cause a logic error when corrupted by an SEU. They also found that two double-bit upsets could cause an IOB to change from an input mode to the output mode. It is important to note that 1) their test bed could not be used to investigate delays induced by IOB configuration bit flips, and that 2) their target FPGA had allowed some I/O standard mixing within a single I/O bank, meaning that the setting of a whole I/O bank could not be changed by single or double bit upsets. Another paper [16] has proposed and applied a methodology for measuring the MBUs cross-section for protons and heavy-ions. Their experimental results showed that Virtex-4 had significantly more IOB events induced by proton irradiation than the previous families (Virtex, Virtex-II, Virtex-II Pro), that IOBs were very sensitive to MBUs induced heavy ion radiation, and that MBUs were on the rise (also reported in [19]). The impact of these MBUs on IOBs behavior was not investigated.

These most recent results are important as they strongly suggest that events affecting IOBs in SRAM-based FPGAs exposed to radiations should become more frequent. As a consequence, more attention should be dedicated to the IOBs. This paper presents results confirming this trend and the relevance of considering SEU events on IOBs. Our results, obtained with a Xilinx Virtex-5 FPGA, show that these events can occur at a rate approaching those occurring in the core routing and CLBs. These results also reveal that very significant changes to path delays can be induced by flipping some bits in IOBs, with the particularity that bit flips must occur in pairs for these delays to appear. Two approaches are combined to show the existence of these delays: 1) emulation with a tool from Xilinx called SEU Controller [2], and 2) proton irradiation at the TRIUMF laboratory [17]. To our knowledge, it is the first time IOBs are the subject of such an investigation where the occurrence and the amplitude of the delay changes affecting them are measured.

## II. EMULATION SETUP

Fig. 1 shows our emulation experimental setup, which comprises a commercial FPGA-based test board (Digilent Genesys [20], containing a Xilinx Virtex-5 XC5VLX50T), a spectrum analyzer (Anritsu Spectrum Master MS2721A), and a personal computer (PC). The Virtex-5 is first used to implement a simple ring oscillator (the ring oscillator inverter is implemented directly in the IOBs, therefore no LUT is used for the design), with an external loop connected to the spectrum analyzer which measures its oscillation frequency. The PC is used to load the configuration bits and to communicate through a USB link with the SEU Controller, which is also implemented in the Virtex-5. The SEU controller is an IP core provided by Xilinx [2] to emulate SEUs within a device by injecting errors in a controlled and predictable way into the configuration memory. Moreover, we developed a complementary software tool in order to ease and automate the emulation.

We implemented the whole design (ring oscillator + SEU controller), coded in VHDL, using the Xilinx regular design flow (ISE environment). Once the configuration load was obtained, we used the Xilinx bitgen command [3] to create a list of the potentially critical bits, saved in an .EBD file. Then we emulated SEUs with the SEU controller using the .EBD file, which contains 186 bits (for the design shown in Fig.1). Our investigations indicated that 90.3% of these bits belong to IOBs while 9.7% are routing bits. Note that not all 186 bits are actually critical. In this paper, a bit is labeled critical when a fault or an observable delay change (ODC) occurs after it is flipped, alone or in combination with others.

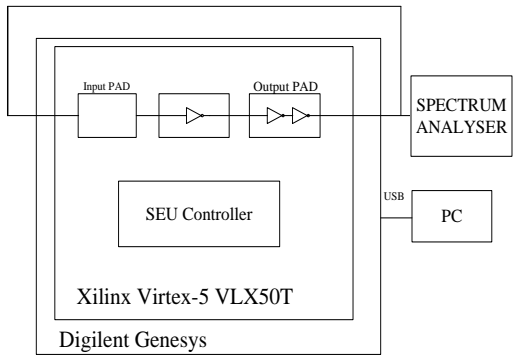

Fig. 1. Emulation setup

### III. EMULATION RESULTS

The emulation was done in two different modes: 1) the SBU mode, where a bit flip is corrected before injecting another one, and 2) the accumulation mode, applied after removing the critical bits detected by the SBU mode, where bit flips are not corrected. No ODCs were observed in the SBU mode. However, bit flips on 5 of the 186 bits caused hard faults (the ring oscillator stopped) in this mode. Interestingly, ODCs were observed while performing the emulation in the accumulation mode. ODCs ranging from 400 to 830ps were measured, as higher or lower values of the nominal 50 MHz ring oscillator frequency. All ODCs were caused by pairs of bit flips on potentially critical bits belonging to the IOBs. This constitutes a key observation as it indicates that each bit involved in such delays is individually not critical, but that they may become critical when other bits are flipped. Further investigations are required to verify that it is always the case, but it is conjectured that this is partly due to a mitigation strategy adopted by Xilinx. A total of 32 emulation experiments were necessary to identify these critical bit pairs. These 32 pairs map to a total of 32 bits from the set of 186 potentially critical bits. Note that one such bit appeared in 9 of the 32 pairs. In addition to these 32 bit flip pairs causing significant observed delay changes, there were 5 other pairs which caused the ring oscillator to stop (even though single bits from these pairs taken in isolation did not). These 5 pairs map to a total of 5 single bits (not critical in the SBU mode).

So far, our investigations have shown that these delay changes are very similar to those obtained when changing the IOB parameters (namely standard, slew rate, output drive [4]) in the User Constraint File (UCF) file before synthesis. The parameter changes were done in two different modes: 1) the singular mode, where only one parameter is changed, 2) the multiple mode, where two or more parameters are changed at the same time. The different changes are applied with respect to the default parameters (standard=LVCMOS25, slew rate= slow and output drive = 12mA). The results of some of the largest expected changes are shown in Table 1. Interestingly, the 2 values (770, 830 ps) reported in this table were also observed while using the SEU Controller.

To our knowledge, it is the first time that such significant delay changes caused by (pairs of) bit flips in SRAM-based FPGA IOBs are reported. These bit flips affecting IOB characteristics can obviously create synchronization problems. This might have a serious impact on embedded systems reliability, for example when FPGAs are used as bridges between different system modules.

Table 1. Absolute value of the largest expected delay changes versus IOB parameters changes; mode changes are highlighted.

| Standard | Slew Rate | Output Drive (mA) | Delay (ps) | RO Frequency (MHz) |
|---|---|---|---|---|
| Singular Change | | | | |
| LVCMOS25 | Fast | 12 | 770 | 52 |
| LVCMOS30 | Slow | 12 | 770 | 52 |
| LVCMOS25 | Slow | 4 | 770 | 52 |
| LVCMOS25 | Slow | 6 | 830 | 48 |
| LVCMOS25 | Slow | 16 | 830 | 48 |
| Multiple Changes | | | | |
| LVCMOS33 | Slow | 2 | 770 | 52 |
| LVCMOS33 | Slow | 4 | 830 | 48 |
| LVCMOS33 | Slow | 8 | 770 | 52 |

### IV. RADIATION TESTING EXPERIMENTAL SETUP

Fig.2 shows the radiation testing setup used for the experiments done at the TRIUMF laboratory. This setup is similar to the one used in our previous experiments at TRIUMF [1]. It comprises a commercial FPGA-based test board (Digilent Genesys, the same used in the emulation setup), a small daughter board (custom PCB) with a 74AC04 chip (with 6 inverters operating at $V_{cc}$= 2.5 V), a spectrum analyzer (Anritsu Spectrum Master MS2721A, the same used in the emulation setup), and a remote PC. This remote PC, used to (re)load and read back the configuration bits, was positioned close to and connected to the test board through a USB link.

Both the spectrum analyzer and remote PC were connected through an Ethernet link to a master PC running a Labview interface we developed for experiment control and data logging. This Labview interface is based on National Instruments' Anritsu MS2721A VISA pilot, which allows remote operation of the spectrum analyzer as well as data logging.

The Virtex-5 was used to implement two ring oscillators (ROs), running at about the same frequencies ($F_1$ and $F_2$, where $F_2 > F_1$). These ROs were created directly in the IOBs using their IOBUF primitives as shown in Fig.2. The first RO occupies 179 IOBs while the second one occupies 160 IOBs (some IOBs are configured as inverters while others are not). Overall, the two ROs occupy 71% of the 480 available IOBs.

Each RO is connected to the input of one inverter (74AC04 chip). The outputs of the two 74AC04 inverters are shorted by a 5.1KΩ resistor, while one inverter output is monitored by the spectrum analyzer. This short allows the generation of a signal with a frequency spectrum containing the difference between the 2 ROs' frequencies, $F_2 - F_1$, where :
- $F_2 \approx 919$ KHz and $F_1 \approx 827$ KHz.
- $F_2 - F_1 \approx 92$ KHz.

## V. EXPERIMENTAL RESULTS

This setup allowed us to perform 3 sets of delay measurement experiments. Each set, comprising 10 experiments, is performed at different energy levels: 63, 50 and 35MeV. Each experiment was stopped when one RO broke. By contrast, a delay change produces a variation in the measured ($F_2 - F_1$) frequency difference. Note that positive and negative frequency difference changes were observed. The frequency difference increase when $F_2$ increases and it decreases when $F_1$ increases. Both RO are exposed to radiations and subject to changes in their oscillation frequency. Figure 3 shows one of the graphs obtained for a delay-fault observation experiment, showing $F_2-F_1$ measurements over time.

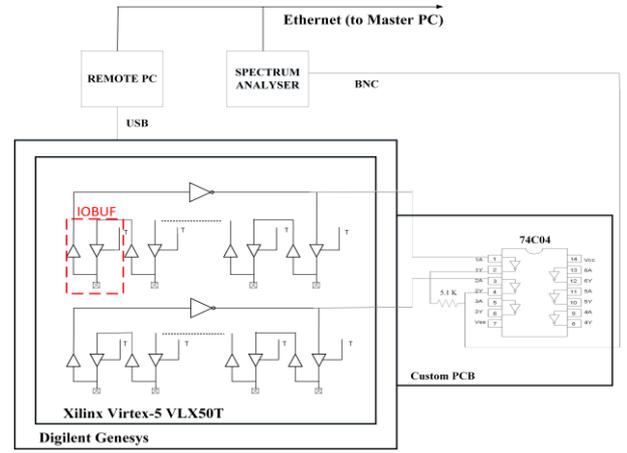

Fig. 2. Experimental Setup

While our experimental results need further analysis in order to extract and validate the ODC magnitude, so far we were able to identify 74 ODCs (the number should slightly increase when our analysis is completed, and will be updated in the final paper). The magnitude of identified ODCs ranges from 63ps to 6.2ns, with an average amplitude of 515ps. If we compare the number of ODCs with the number of broken ROs (BROs) we obtained during our 30 experiments, it leads to a ratio of the number of ODCs to the number BROs of about 2.5, which is a clear indication that ODCs occur more often than BROs. We are currently investigating the effects of these delays on a real design. The results of this investigation will be presented in the final version of this paper.

We were also interested by the IOB sensitivity with respect to the other FPGA generic resources (namely CLB + routing). Let us define the total number of IOB events as the number of ODCs and BROs observed during the 30 delay-fault observation experiments. Let us also define the IOB event cross-section, IOB_CS, as the total number of IOB events divided by the overall fluence for the 30 delay experiments. According to our estimation, we obtained a total IOB event cross-section of 6.2e-10 $cm^2$, which covers 339 of the 480 available IOBs. If we extrapolate to all available IOBs, we obtain an extrapolated IOB event cross-section, E_IOB_CS, of 8.7e-10 $cm^2$. Similarly to the IOB events, let us define the total number of CLBR (CLB + routing) events as the number of ODCs and BROs observed during (42) other delay experiments (also performed at 63, 50 and 35MeV) similar to the one of [1], with 2 ROs occupying 50% of the FPGA slices. Let us also define the CLBR event cross-section, CLBR _CS, as the total number of CLBR events divided by the overall fluence for the 42 related delay experiments, for which we got an estimated value of 8.5e-10 cm$^2$. If we extrapolate to all available slices, we obtain an extrapolated CLBR event cross-section, E_CLBR_CS, of 1.7e-9 cm$^2$.

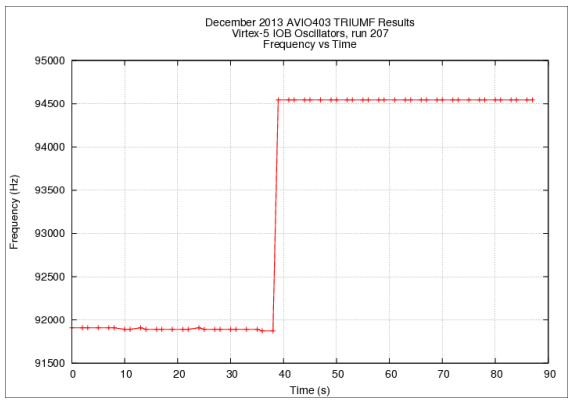

Fig. 3. Experimental results, run 207; ODC of 3.9ns.

The E_CLBR_CS / E_ IOB_CS ratio, whose value is slightly lower than 2, indicates that on average, 33% of the events affecting the target Virtex-5 generic resources (IOBs, CLBs, routing) occur in the IOBs.

As mentioned before, the maximum magnitude of a single ODC that was observed is 6.2ns. In fact, 4 ODC events had a value larger than 1.5ns. We conjecture that such large single ODC event values were caused by an accumulation of delays along the RO within a same I/O bank. Indeed, in a Virtex-5, all the IOBs belonging to an I/O bank must share the same I/O standard. Thus, it is likely that the IOBs also share the same configuration bits defining the standard. If this assumption is correct, several outputs of a same bank can have their timing characteristics simultaneously modified. One possible consequence is to make mitigation strategies such as Triple Modular Redundancy (TMR) inefficient to tolerate IOB delay faults if the 3 outputs of a same TMR group are affected the same way at the same time.

## VI. CONCLUSION

In summary, we developed an experimental setup to emulate bit flips caused by radiations on a Virtex-5 FPGA using the Xilinx SEU Controller. The emulation experiments we performed revealed that bit flips on configuration bits controlling IOBs could lead to significant delay changes as large as 830ps, and that these delay changes (as well as some functional errors) were induced by pairs of bit flips. We also presented results obtained at the TRIUMF laboratory that confirmed the existence of these delays in IOBs. In addition to confirming our results obtained by emulation, our experiments also showed that much larger delay changes up to 6.2 ns could be observed. Combining these results with those of other experiments allowed showing that faults in IOBs can represent a significant part of all faults affecting the generic resources of an FPGA.